# Effect of Strain Rate on Single Tau, Dimerized Tau and Tau-Microtubule Interface: a Molecular Dynamics Simulation Study

Md Ishak Khan[1], Kathleen Gilpin[2], Fuad Hasan[1], Khandakar Abu Hasan Al Mahmud[1] and Ashfaq Adnan[1, +]

[1]Department of Mechanical and Aerospace Engineering, University of Texas at Arlington, Arlington, Texas, 76019, USA

[2]Academic Partnership and Engagement Experiment (APEX), Wright State Applied Research Corporation, Beavercreek, OH, USA

[+]Corresponding Author. Email: aadnan@uta.edu

**Abstract**

Microtubule-associated protein (MAP) tau is a cross-linking molecule that provides structural stability to axonal microtubules (MT). It is considered as a potential biomarker for Alzheimer's disease (AD), dementia and other neurological disorders. It is also a signature protein for Traumatic Brain Injury (TBI) assessment. In the case of TBI, extreme dynamic mechanical energies can be felt by the axonal cytoskeletal members. As such, fundamental understandings of the responses of single tau protein, polymerized tau protein and tau-microtubule interfaces under high-rate mechanical forces are important. This study attempts to determine high-strain rate mechanical behavior of single tau, dimerized tau and tau-MT interface using molecular dynamics (MD) simulation. The results show that a single tau protein is a highly stretchable soft polymer. During deformation, first, it significantly unfolds against van der Waals and electrostatic bonds. Then it stretches against strong covalent bonds. We found that tau acts as a viscoelastic material, and its stiffness increases with the strain rate. The unfolding stiffness can be ~50-500MPa, while pure stretching stiffness can be >2GPa. The dimerized tau model exhibits similar behavior under similar strain rates, and tau sliding from another tau is not observed until it is stretched to >7 times of original length, depending on the strain rate. The tau-MT interface simulations show that very high stretching is required to separate tau from MT suggesting Tau-MT bonding is stronger than MT subunit bonding between themselves. The dimerized tau-MT interface simulations suggest that tau-tau bonding is stronger than tau-MT bonding. In summary, this study focuses on the structural response of individual cytoskeletal components, namely microtubule (MT) and tau protein.

Furthermore, we consider not only the individual response of a component, but also their interaction with each other (such as tau with tau or tau with MT). This study will eventually pave the way to build a bottom-up multiscale brain model and analyze TBI in a more comprehensive manner.

1. **Introduction**

Blast-induced TBI has been considered highly significant on the battlefield and in sports (Okie, 2005; Taylor and Ford, 2009). Numerous studies have been performed to address this injury mechanism and consequences from different perspectives, such as post-traumatic stress disorder, aggregation of neurological disorders, damage threshold of cytoskeletal components, etc. When head is impacted by blast-like mechanical force, the resulting pressure wave transmits to the interior of the brain and causes macroscale, tissue-level, and cellular level damage (Al Mahmud et al., 2020; Hasan et al., 2021; Wu and Adnan, 2018). A major part of brain's interior is built on over 100 billion of neuron cells and its surrounding extra-cellular matrices. Structurally, a single neuron is composed of a soft cell body (soma), fibrous dendrite branches, and an axon fiber. Both axons and dendrites are supported by the cytoskeleton, which is primarily composed of neurofilaments (NF), microfilaments (MF), tau proteins and microtubules (MT). Of these, the mechanical behavior of MF and MT have been studied in depth, and of NF has been studied partially in the past. However, mechanical response of tau protein and its interaction with MT have not been studied in detail. One reason is the structural disorder in tau protein. It is known as an intrinsically disordered protein (IDP) implying a definitive structural conformation of this protein is unavailable. As such, fundamental understandings of the responses of single tau protein, polymerized tau protein and tau-microtubule interfaces under high-rate mechanical forces are very important. Yet, comprehensive mechanical behavior of tau protein and tau-MT interaction are not available in the literature.

As tau protein is an intrinsically disordered protein (IDP), a brief introduction on IDP is relevant for this study. Intrinsically disordered proteins (IDP) are special proteins involved in many different cell-signaling pathways within the cell. They have unique capabilities of performing different functions based on conformations that occur due to different post-translational modifications, different binding substrates (proteins, nucleic acids, fibers, etc.) and fewer of the order-promoting amino acids found within hydrophobic cores of proteins. These include tryptophan, cysteine, tyrosine, leucine, phenylalanine, isoleucine, and valine. On the other hand, the amino acid make-up of an IDP commonly includes an abundance of amino acids associated with disorder, including alanine, arginine, glycine, glutamine, serine, proline, glutamate, and lysine (Jorda et al., 2010). In recent years, IDPs have been more prominent in biomedical research in an effort to understand their variable roles. This paper discusses tau, which is heavily involved in Alzheimer's Disease (AD), but Parkinson's disease (alpha-synuclein), Amyotrophic Lateral Sclerosis (superoxide dismutase-1), and Huntington's Disease (poly-glutamine gene products) are also caused by IDPs (Uversky, 2010). Without stable conformations and singular, defined functions, IDPs are prone to aggregation and once aggregated, burden the cell's degradation machinery.

After clarification on IDPs, we can move on to the specific introduction on their deformation characteristics and mechanical behavior. Some recent works seek to obtain mechanical properties of MT or analyze the mechanism of damage and failure of axonal microtubules (Wells and Aksimentiev, 2010; Wu and Adnan, 2018). The structural unit of MT (the αβ-tubulin) has been determined earlier by electron crystallography (Nogales et al., 1998). Further studies have shown that there are two distinct sites at the C-terminal on each subunit, to which tau protein can be attached. One of them is the C terminal site, which contains the C-terminal 12 amino acids of MT, and another one is the internal site, which is situated at the last 1-3$^{rd}$ of C-terminal excluding the last 12 amino acids. Tau interacts with the C-terminal site by R1 or R1-R2 interrepeat, and with the internal site by the rest of the MT binding sites (R2-R4). It has also been proposed that tau-MT interaction can take place in two manners: i. tau may interact with only α or β subunit of an MT by the mechanism above or ii. tau may interact with both portions of α and β subunits (one with R1 or R1-R2 interrepeat, another with R2-R4) (Chau et al., 1998). Further studies on localization of tubulin binding sites for tau protein (Maccioni et al., 1988; SERRANO et al., 1985) strengthened the finding of Chau et al. As important is MT as an individual cytoskeletal component, so are the crosslinks between them, which are denoted by microtubule associated proteins (MAP), mainly tau. Tau is an intrinsically disordered protein (IDP), and has no defined secondary structure (Rosenberg et al., 2008). However, there are numerous predictor software for obtaining secondary structures of proteins, such as i-TASSER (Zhang, 2008), Phyre2 (Kelley et al., 2015), etc. which can predict the structures of IDPs with a certain level of confidence. The details of the mechanism of predicting secondary structures is out of the scope of this manuscript, although relevant discussion is presented in the supplementary material.

Numerous studies regarding tau specifically discuss on phosphorylation and hyperphosphorylation, which are post-translational modifications. There have been separate studies on tau structure, which suggest that phosphorylation level plays a critical role to distinguish healthy tau from pathological tau (Becker and Przybylski, 2007), and it also has been hypothesized to alter the ability of tau to bind MTs as well as other functionalities (Hanger et al., 2009). Earlier Monte Carlo simulation study that has attempted to find out the threshold of tau pathology suggested that numerous candidate amino acids can be phosphorylated, but under pathological conditions around 7 of them are actually phosphorylated (Jho et al., 2010). The abnormally phosphorylated tau proteins are 3-4 times more phosphorylated than the normal ones, and ~2 sites per mole of tau protein can be phosphorylated in normal condition (Kenessey and Yen, 1993). Recently, cryo-EM technology has been able to obtain 3-4Å resolution image of paired helical filaments (PHF) in AD affected brains (Fitzpatrick et al., 2017), which is directly related to abnormal phosphorylation in tau.

Not only phosphorylation and dephosphorylation are relevant, but also the strain rate dependent structural response of tau regarding this study. In some simulation studies of microtubule bundles cross-linked with tau protein, the response under different levels of uniaxial tension for different distributions of cross-links and discontinuities have been observed, and the estimated Young's modulus used for calculation was 5MPa (Peter and Mofrad, 2012), although they have admitted that it is decidedly approximate, and tau-tau, tau-MT, etc. interactions are important candidates to determine the mechanical behavior. Their model showed that bundle failure occurred due to failure of cross-links. Tau proteins, through complementary dimerization with other tau proteins, form bridges to nearby microtubules to form bundles (Mandelkow et al., 1995). The tau proteins have edge-to-edge distance of 20nm and hexagonal packing (Chen et al., 1992). Furthermore, computational models have been built to study the behavior of cytoskeletal filaments (Chandran and Mofrad, 2009; Mofrad, 2008) and cross-linked networks have also been investigated (Claessens et al., 2006; Kim et al., 2009; Silber et al., 2004). Earlier computational and theoretical models have shown that shear resistance provided by the cross-linked network greatly increases MT bending stiffness (Tolomeo and Holley, 1997), and found bundle stiffness regime (Bathe et al., 2008). Discrete bead-spring models have been widely used to build filament and network structure (Rodney et al., 2005; Sandersius and Newman, 2008). In another finite element analysis, the Young's Modulus of tau protein was assumed to be 5MPa (Shahinnejad et al., 2013), but elevated to 62.5 MPa in order to reduce the flexural behavior so that the spring constant can be the same as Peter et al (Peter and Mofrad, 2012). The tensile test results reasonably agree with that of Peter et al., and it also analyzed the behavior of bundle under torsion. Prior continuum and computational axon modeling gives us useful insights on tau protein, such as viscoelastic shear lag model (Ahmadzadeh et al., 2014), failure mechanism of axon study (de Rooij and Kuhl,

2018), and so on. The computational continuum models and rate-dependent tests can determine the properties of the cytoskeletal components such as microtubules, and therefore are relevant to the current study.

The structure and phosphorylation studies cannot depict the complete structural and functional scenario of tau, because another critical aspect of tau protein is the MT binding region, which can work as a backbone of the structure and from which the projection domain can spread around (Lee et al., 1989). Some recent studies have further clarified the position and structure of MT binding sites in tau, such as Rosenberg et al (Rosenberg et al., 2008). Each of the four binding regions of tau is around 18 amino acids (AA) long, separated by interrepeats of 13-14 AA length. The positively charged repeat or interrepeat region is believed to facilitate electrostatic interactions between tau and negatively charged surface of MT (Chau et al., 1998; SERRANO et al., 1985; Silber et al., 2004). In the N-terminal of the MT binding region, there is a positively charged and proline-rich region, which has phosphorylation sites responsible for the regulation of MT association to taus (Goode et al., 1997; Trinczek et al., 1995). The other portions of tau protein have significant functionalities as well, such as the N-terminal region, which can contain zero to two negatively charged inserts (each 29AA long). The N terminus and proline-rich region make up the projection domain of tau protein, which extends outward from the MT surface and determines inter-MT distance in MT bundles (Chen et al., 1992). Other studies show that MT binding with tau is very fast, does not depend on the MT location in the axonal shaft and varies with MT curvature (Samsonov et al., 2004). Study of intrinsically disordered tau protein folding on MT shows that tau locally folds into a stable structure upon binding (Trinczek et al., 1995).

Aside from studying standalone tau protein response, our interest is to obtain insight on tau-MT interaction. Tau protein organization on MT can be described as a coating or surface decoration, as suggested by some works (Santarella et al., 2004). The mode of tau protein binding with MT has been studied, and it suggests that tau protein binds along as well as across protofilaments (Al-Bassam et al., 2002; Kellogg et al., 2018).

Now, while considering a feasible approach to computationally determine single tau, dimerized tau, and tau-MT behavior, we must consider the earlier studies that have been performed on intrinsically disordered proteins (IDPs). Therefore, by focusing on computational studies on IDPs (especially on tau proteins), we have found that recent years have observed significant improvement in computational studies by using predicted structure. Notable examples are electrostatic study (Castro et al., 2019). and aggregation behavior study on tau (Battisti et al., 2012). However, studies focusing on mechanical response of tau are still non-existent. Recently published comprehensive review study has pointed out this current ongoing limitation (Khan et al., 2020a), and an MD simulation study has shown the effect of phosphorylation on tau (Khan et al., 2020b). Therefore, this study is an attempt to address one of the significant ongoing limitations in the current literature by applying strictly mechanical loading.

Based on the existing literature as discussed above, there are certain parameters yet undetermined, such as mechanical behavior single tau filament due to strict mechanical loading (deformation at high strain rate), required stretching for separation of dimerized tau proteins, tau-MT binding strength and whether MT subunit binding between themselves is stronger than tau-MT bond, etc. This study attempts to address these specific questions using atomistic computational method called molecular dynamics (MD). The rationale behind choosing this tool are twofold: 1) reliable experimentation at this length scale is rare and 2) MD is an appropriate tool to capture the molecular details of the elements involved. Specifically, for the single tau, we have obtained the secondary structure from i-TASSER predictor software (Zhang, 2008). For dimerized tau, we have made the model with overlapped projection domain as depicted by Rosenberg et al (Rosenberg et al., 2008). For tau-MT interaction, we have used Chau et al proposals of interaction, where tau can interact with one or both subunits of MT (Chau et al., 1998). The effects of high strain rate on the deformation mechanism of single tau, dimerized tau and tau-MT interface are studied.

2. Method

The tau protein structure is obtained from the i-TASSER predictor software (Zhang, 2008). We have used the model with the C score of 1.06 (the C score is determined based on significance of threading template alignments and the convergence parameters of the structure assembly simulations) which we have assumed satisfactory for an IDP. For single tau protein, we have solvated the obtained structure with TIP3P water using CHARMM-GUI (Jo et al., 2008) solvator and quick-MD simulator modules. Required number of 0.15M KCl ions were added to obtain charge neutralization for explicit solvent simulations. The dimerized model was created by using UCSF Chimera (Pettersen et al., 2004), in which we have overlapped the projection domains of two identical tau proteins. We have used implicit solvent technique of CHARMM (Best et al., 2012) in LAMMPS (Plimpton, 1995) (pair_style lj/charmm/coul/charmm/implicit command) for dimerized tau, which facilitates a bigger box size, faster calculation and convenient observation of two tau system. The implicit solvent computes with a modification of adding an extra $r^{-1}$ term for Coulombic energy calculation and skipping long range Coulombic energy calculations.

The MT structure is obtained from the existing model built by Wells et al (Wells and Aksimentiev, 2010), each subunit of which has one GTP or a GDP, along with one $Mg^{2+}$ ion in the junction. By repeating the helically arranged 13 subunits periodically along the length direction, we have created a virtually infinite MT (similar methodology has been followed in the work of Wu et al for MT study (Wu and Adnan, 2018)). The tau-MT interaction system was created by placing the tau binding sites in close proximity to the MT binding sites as proposed by Chau et al (Chau et al., 1998). We have used the implicit solvent technique for tau-MT as described above for dimerized tau.

With periodic boundary conditions in all three directions, we have minimized and equilibrated the structures for all cases (single tau, dimerized tau and tau-MT) for 100ps to minimize the potential energy at a targeted temperature of 310K. Single tau, dimerized tau and tau-MT contained 6424, 12848 and 33292 atoms respectively. The LJ potentials are used with inner and outer cutoff of 10Å and 12Å, respectively. Long range coulombic interactions are computed by pppm style, facilitating a particle-particle particle-mesh solver with a 3d mesh.

We have used CHARMM36 (Best et al., 2012; Brooks et al., 2009; MacKerell et al., 1998) potential parameters with appropriate CMAP corrections (Best et al., 2012) for all the simulations. Potentials for GDP and GTP are taken from that of ADP and ATP respectively (Wells and Aksimentiev, 2010). The equilibration was performed in NVT canonical ensemble, with the temperature damping parameter of 100fs.

In single tau protein tensile tests, we have fixed the MT binding domain and pulled the projection domain at different strain rates ($10^8$-$2x10^9$ $s^{-1}$) along x axis. For dimerized tau, we have fixed the MT binding domain of one protein and pulled away another binding domain along x axis ($10^9$ $s^{-1}$ and $2x10^9$ $s^{-1}$). For tau-MT, While keeping the upper and lower few layers of atoms fixed of the MT, the tau protein was pulled away by few atoms of the projection domain (strain rate: $10^9$ $s^{-1}$ and $2x10^9$ $s^{-1}$). For the relevance of the simulations with explicit water molecules as solvent vs implicit water molecules, readers may refer to the supplementary materials of Wu et al (Wu and Adnan, 2018). Although biomolecule simulations are more realistic with explicit solvation, we have adopted implicit technique for dimerized tau and tau-MT due to the high box size required for the high stretch, and for convenient observation of unfolding, stretching and separation.

The stress-strain plots are obtained by the per-atom stress calculation and summation in LAMMPS. However, as the output is in (pressure x volume) unit, we must divide the obtained stress value by the volume of the protein (or certain portion of the protein). The general formulation used by stress per atom command is $P = (P_{xx}+P_{yy}+P_{zz})/(3xV)$, where

$P_{xx}$, $P_{yy}$ and $P_{zz}$ are the summations of stress/atom values for all atoms in x, y, and z direction respectively, and V is the summation of volume of the atoms of the protein being considered. The approximated volume was obtained by Voronoi cell approximation, adapted from LAMMPS voro++ package (Rycroft, 2009). The strain is simply obtained by the displacement of the atoms from the initial position. All the tensile tests are performed in NVT ensemble, with 100fs temperature damping parameter. The visualizations of the tensile tests are carried out by OVITO software (Stukowski, 2010).

All of the simulations were carried out by the STAMPEDE2 supercomputer of Texas Advanced Computing Center (TACC).

## 3. Results

### 3.1 Single Tau Deformation

For the single tau model, we have performed the tensile tests at four different strain rates: $10^8$, $5 \times 10^8$, $10^9$ and $2 \times 10^9$ $s^{-1}$. The MT binding region atoms are fixed, and first few atoms of the projection domain are pulled at –x direction. The calculated stress-strain graphs are shown in Fig. 1.

It can be observed that the stress-strain response of a single tau can be divided into two regions - the "unfolding" and "stretching" regions. When the mechanical load is applied, first the folds disappear one after another because of breaking of the electrostatic and van der Waals forces between the folded portions of the filament (this phenomenon will be called as "unfolding" onwards in this manuscript). However, the mentioned forces are relatively weaker than chemical bonds, as they are merely the result of proximity and positioning (3D confirmation of the structural portions in space). Therefore, during the unfolding region (see Fig. 1) no significant rise of stiffness is observed. However, after the filament is free of the folded regions, the structurally linear filament will be deformed against the relatively stronger covalent bonds (this phenomenon will be called as "stretching" onwards in this manuscript). Therefore, in this region there will be higher stress developed and significant rise in stiffness will be observed, as shown in the "stretching region" in Fig. 1. The unfolding and stretching phenomena are clarified by simplified schematic in Fig. 2 as well.

In Fig. 1, multiple distinct slopes can be observed that reflect the stiffnesses of tau protein at different strain-states. Up to ~200% tensile strain, the slope is very small implying a relatively low unfolding stiffness of tau. After the unfolding is complete, we can observe pure stretching of the covalent bonds. Unfolding stiffness obtained from the strain of 0%-150%, while stretching stiffness is obtained from the strain of 200%-300% where another distinct change of slope is observed. Fig. 2 shows the unfolding and stretching snapshots. The unfolding stiffness and pure stretching stiffness values are shown in Table 1. According to Table 1 and Fig. 1, it can be inferred that a single tau exhibits viscoelastic behavior to some extent, and that its stiffness is strongly dependent on strain rate. With the increment of strain rate, tau acts as a stiffer material in both the unfolding and stretching zone, which is expected for viscoelastic soft biomaterial. The slope is in the Mega Pascal range: 0.5GPa or ~500MPa. From the existent literature we have not found any validated estimate of the stiffness of the projection domain of tau, but we have detail information regarding the stiffness of other cytoskeletal components, such as MT and microfilaments. Deformation studies and viscoelastic shear lag models of MT show that the MT stiffness can be in the giga pascal range (Ahmadzadeh et al., 2014; Peter and Mofrad, 2012). Studies on microfilament show that their stiffness can vary from few mega pascals to >2GPa (Higuchi et al., 1995; Huxley et al., 1994; Kim et al., 2015; Kojima et al., 1994; Wakabayashi et al., 1994). Therefore, we can assume that our results obtained for the tau protein stiffness, both in unfolding and stretching region, are reasonable.

**Table 1:** Unfolding stiffness and stretching stiffness of the projection domain of single tau

| Strain Rate ($s^{-1}$) | Unfolding Stiffness (MPa) | Stretching Stiffness (GPa) |
|---|---|---|
| $10^8$ | 50.1 | 0.076 |
| $5 \times 10^8$ | 129 | 0.549 |
| $10^9$ | 284.9 | 1.198 |
| $2 \times 10^9$ | 500.5 | 2.399 |

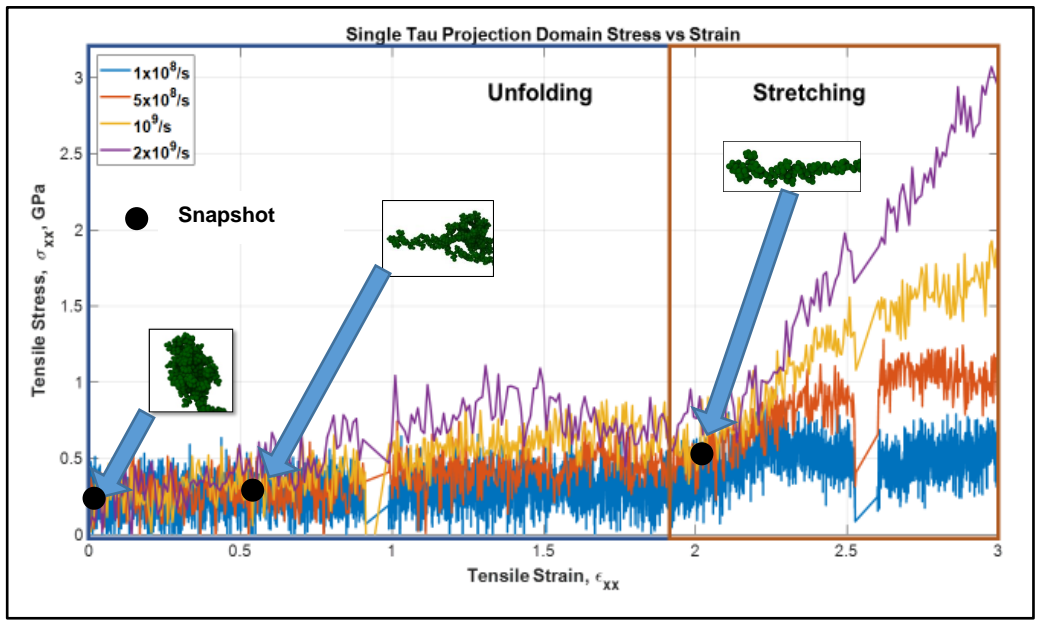

**Figure 1:** Stress vs strain plot of single tau projection domain. Up to ~200% strain, the protein keeps unfolding, and after that a sharp rise in the slope is observed, suggesting the pure stretching of covalent bonds. Unfolding and stretching zones are shown by rectangular boxes.

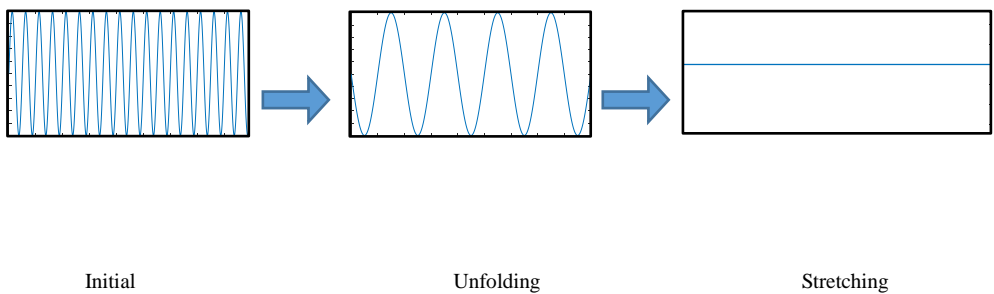

Initial     Unfolding     Stretching

**a. Simplified schematic for protein unfolding and stretching**

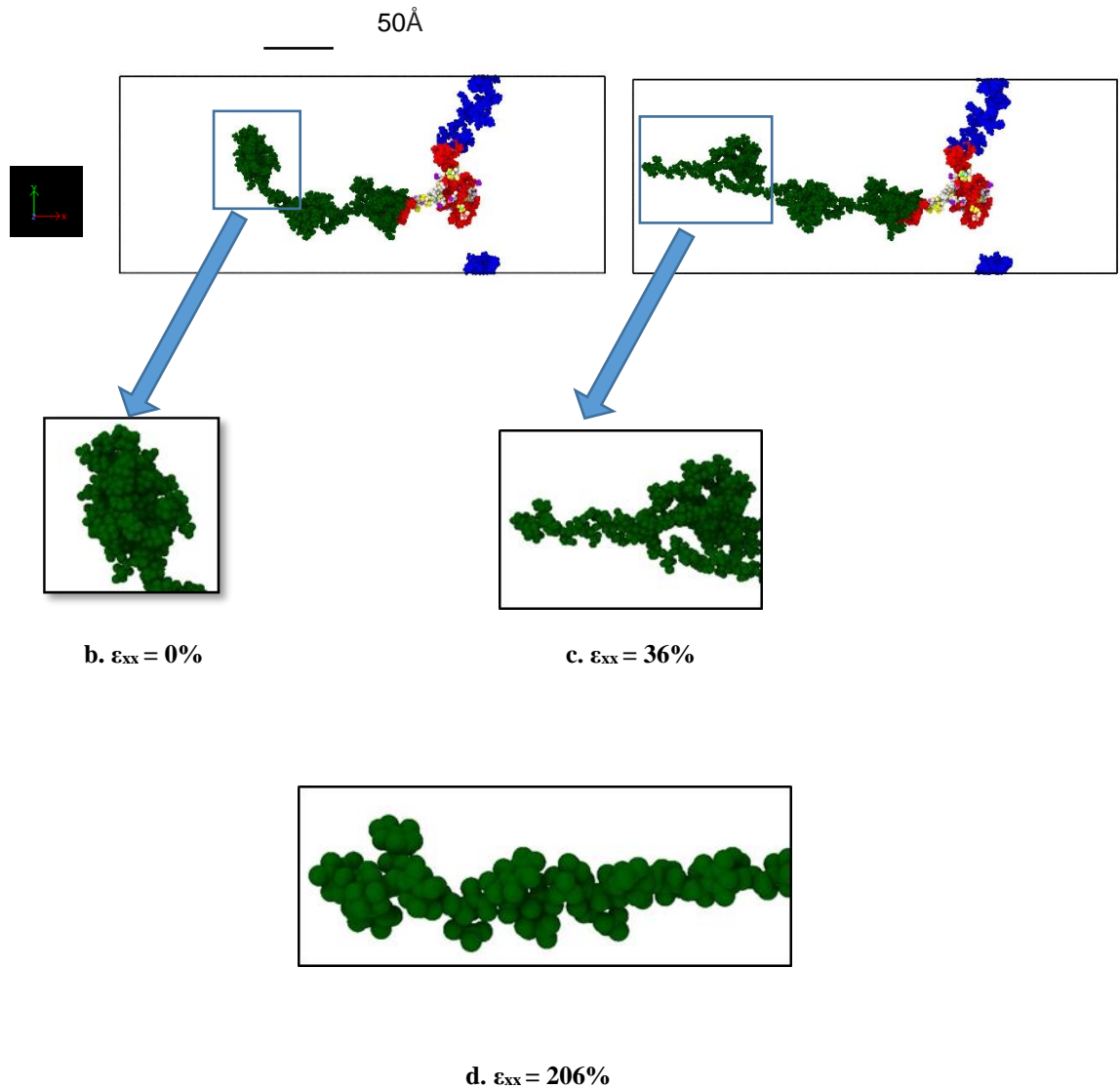

**Figure 2:** a. Simplified schematic for tau protein unfolding and stretching. Initially, there are multiple folding (which are distinct from each other). When loading is applied, the van der Waals force and electrostatic force between the folded portions are broken, and therefore the structure unfolds. At extreme strain, the structure becomes free of all the folds, and stretch to a relatively linear filament. b. Initial single tau structure (strain = 0%), c. tau protein being unfolded due to pulling at $10^9$ s$^{-1}$ (strain = 36%), d. tau protein being stretched (strain = 206%, only the projection domain is shown for the convenience of visualization). Color legends: green: projection domain, red: MT binding region, blue: N terminus or tail, white and the rest: interrepeats between the MT binding regions. The enlarged snapshots are for the first ~1100 atoms for convenient visualization. Water molecules are not shown.

## 3.2 Dimerized Tau Deformation

The dimerized tau model was similar to the single tau, except that we have fixed the MT binding site of one tau and pulled away the binding site of another tau to observe the developed stress and possible sliding out of tau projection domain at extreme strain. We have plotted the stress-strain curves (calculation procedure was similar to single tau) for the projection domain of the tau protein that has been pulled at the x direction, at the strain rate of $10^9$ s$^{-1}$ and $2 \times 10^9$ s$^{-1}$. The implicit-solvent dimerized tau model shows several stages of tension during the test. We are referring to the protein with fixed MT binding region as protein 1, and the protein being pulled as protein 2. The stages observed are (for the strain rate of $2 \times 10^9$ s$^{-1}$): i. Unfolding of protein 2 (up to 163% strain), ii. Stretching of protein 2 (up to 257% strain), iii. Unfolding of protein 1 (up to 334% strain), iv. Stretching of protein 1 (up to 395% strain), v. Disentanglement of the overlapped projection domains of the tau proteins (up to 721% strain), vi. Sliding out or projection domain along with stretching (very fast, occurs at around 722% to 758% strain region), and vii. Separation of proteins (~758% strain). For strain rate of $10^9$ s$^{-1}$, the separation occurs just above the stretch of ~800%. Therefore, the observations imply that the separation of dimer also depends on the strain rate, and at a certain high strain rate, the separation occurs significantly early, although the stretch following the unfolding seems the same for both strain rates. It is to be noted that the separation stretch that are mentioned here are the percentage of strain at which the two dimers are visibly separated from each other, although the pulling away of one tau from another starts earlier (around 30% of strain earlier), and before complete separation there are sub-stages of untangling of the overlapped projection domains of the two tau proteins.

Fig. 3a shows the stress-strain curves of the projection domain of protein 2 and Fig. 4 shows snapshots of the stages observed. The single tau study on this paper already shows that tau is highly stretchable. Therefore, when we are considering dimerized protein model, it is expected for both proteins to be highly stretched before the observation of sliding out of one projection domain out of another. This set of simulation provides new insight about shear mechanism and sliding threshold of dimerized tau, which is yet non-existent in literature. Fig. 3b shows the potential energy trend, and how the potential energy decreases drastically at the separation, irrespective of applied strain rate.

a
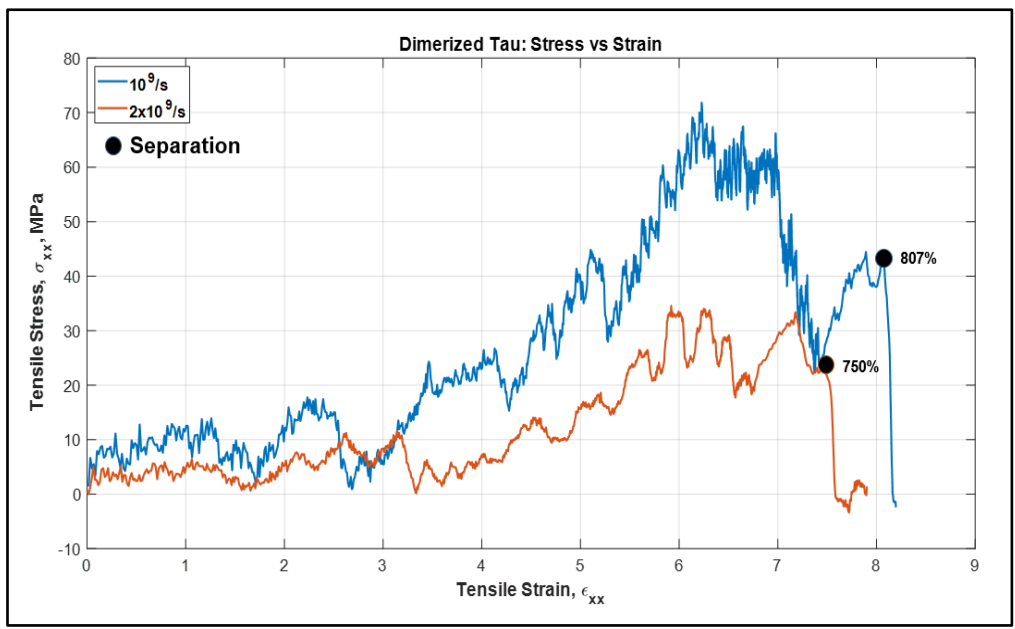

b
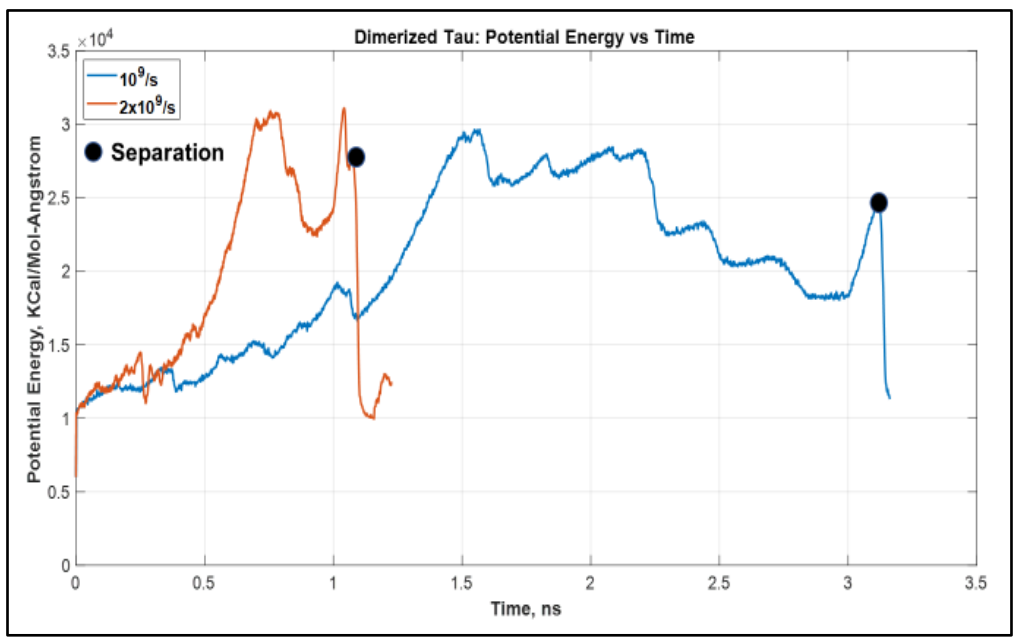

**Figure 3:** a. Stress-strain plot of protein 2 projection domain for two different strain rates. At lower strain rate, we observe development of stress at higher value and delayed separation, and vice versa. b. Potential energy vs time plot for the system.

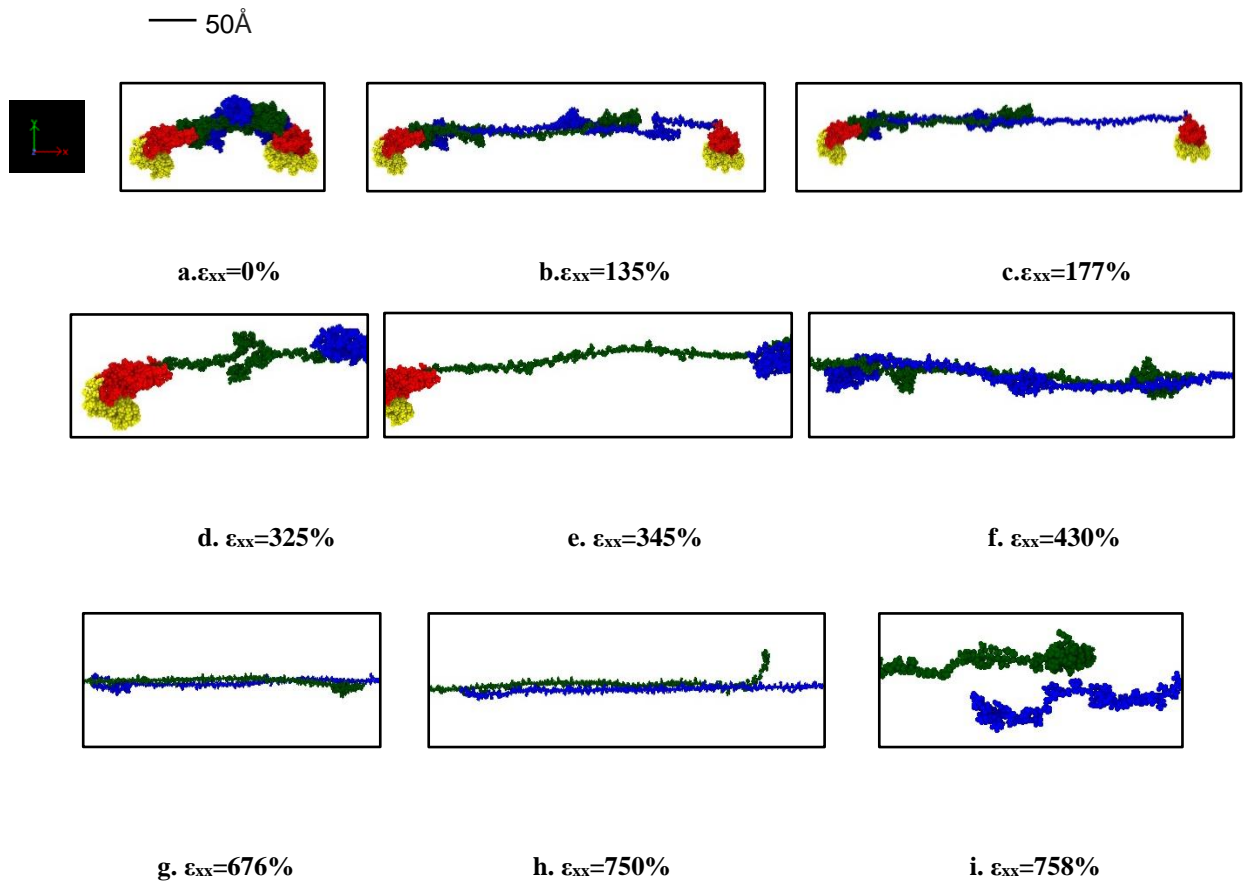

**Figure 4:** Stages observed during the pull of one protein in the dimerized tau model (for the strain rate of $2 \times 10^9$ s$^{-1}$). a. Initial stage (strain: 0%), b. Unfolding of protein 2 (strain: 135%), c. Stretching of protein 2 (strain: 177%), d. Unfolding of protein 1 (strain: 325%), e. stretching of protein 1 (strain: 345%), f. Disentanglement of the overlapped projection domains of the tau proteins (strain: 430%), g. Continued disentanglement (strain: 676%), h. Sliding out of projection domain (strain: 750%), i. Separation of proteins (strain: 758%). Color legends: Green: Projection domain of protein 2, Blue: projection domain of protein 1, Red: MT binding region (including the interrepeats) for protein 1 and 2, Yellow: N terminal tails of protein 1 and 2.

## 3.3 Tau-MT Interaction

In order to determine the nature of single tau-MT interaction, one N-system MT adopted from Wells et al (Wells and Aksimentiev, 2010) was used. One tau protein MT binding site was attached to the surface of the MT. The tensile test results were quite straight-forward: the stretching is followed by unfolding in the projection domain of the tau structure, and even at these extreme strain rates, tau had to be stretched for >11 times of its initial length before getting completely separated from MT. The unfolding and stretching manners are similar to the single and dimerized tau tests, that is, projection domain gets unfolded for a long time, then gets significantly stretched, leading to eventual separation from the MT surface. As we have seen in case of the dimerized tau models, we observe earlier separation at higher strain rate and vice versa for tau-MT models. We can conclude from the two strain rate results that at a certain strain rate range, the tau can be stretched significantly while still being attached to the MT surface. On the other hand, at higher strain rate, it is not allowed to be stretched in that manner, depicting the effect of strain rate. In the dimerized tau, we have observed development of higher stress in tau before final separation for lower strain rate. However, in tau-MT interaction, we have not observed any significant difference in the stress-strain trend for the two different strain rates, rather only in the separation stretch. The untangling sub-stages also take place before separation, as we have observed in dimerized tau. Fig. 5a shows stress-strain graphs for the applied strain rates, and Fig. 5b shows the potential energy graph, which suggests that potential energy decreases significantly at separation, for both cases. Fig. 6 shows various significant stages during pulling of single tau away from MT surface.

Lastly, we are also interested to find out the strain rate dependent nature of dimerized tau-MT interaction. We have attached a dimerized tau on the surface of MT in a similar way of single tau-MT model, and applied high strain rate. This simulation was to compare the relative strength of tau-MT bond to tau-tau bond, and as the simulation shows, tau-tau bond is much stronger than tau-MT bond, because although the pulled tau was stretched significantly at the higher strain rate (~360% before separation from MT surface), it did not disentangle from the other tau. Rather, the entire dimerized structure got separated from MT surface, suggesting that tau-tau bond is stronger than tau-MT bond. For the lower strain rate, the dimerized tau subunits got significant sliding over each other, but eventually got separated from the MT surface before getting entirely separated from each other (~825% before separation). Fig. 7 shows that potential energy significantly reduces at separation, and Fig. 8 shows different significant stages up to and at separation.

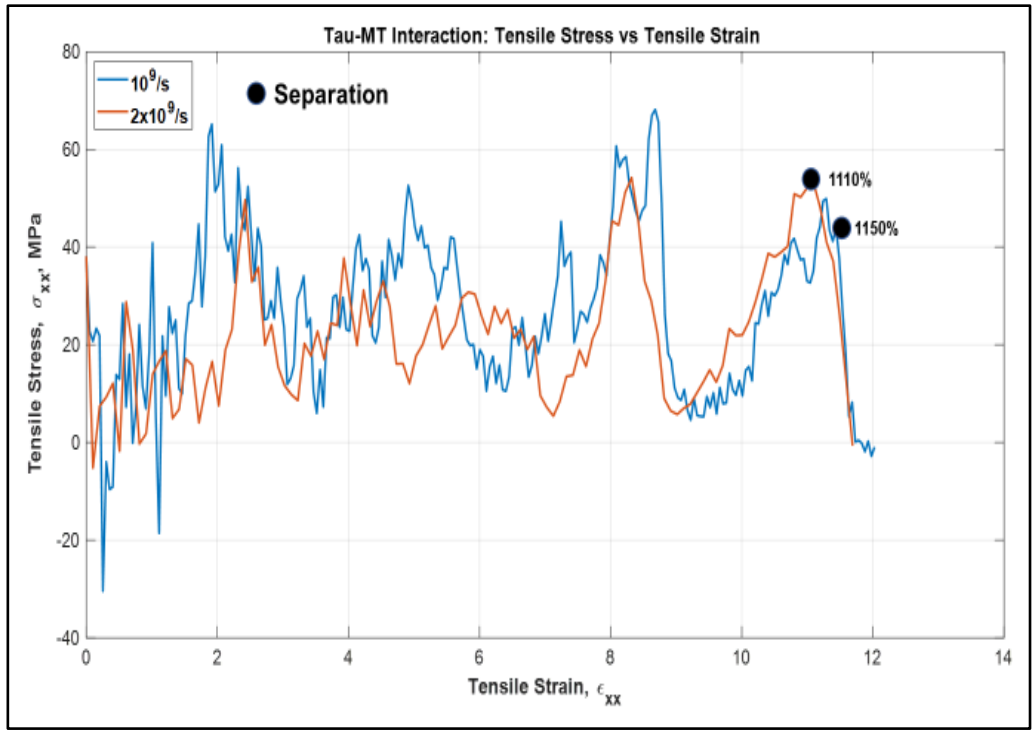

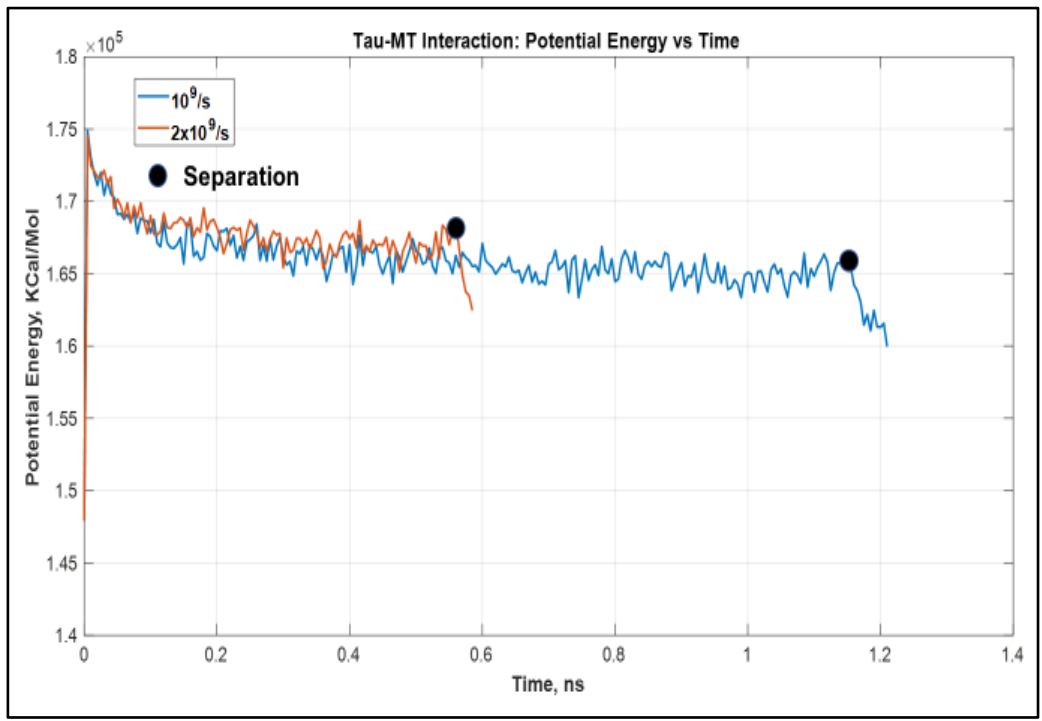

**Figure 5:** a. Stress vs Strain graph for the projection domain of tau during the pulling at different strain rates. The stress-strain trends are similar for both the strain rates, although the separation occurs at different strain. b. Potential energy vs time plot for the single tau-MT system.

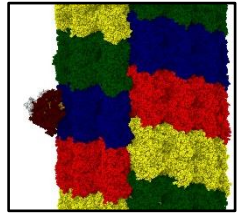
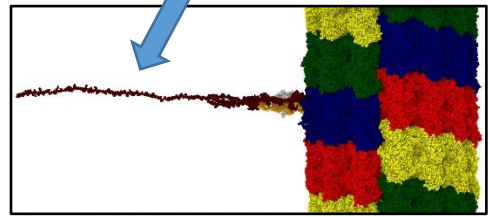

**a.** $\varepsilon_{xx}=0\%$  **b.** $\varepsilon_{xx}=409\%$

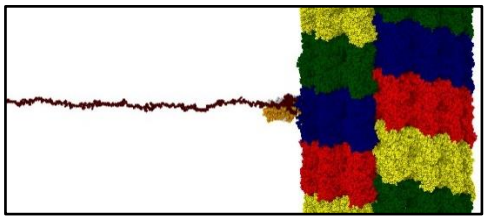
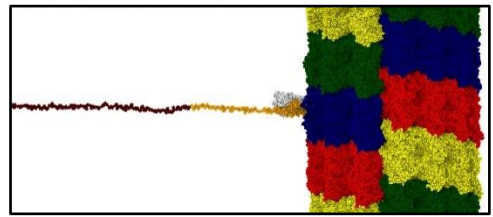

**c.** $\varepsilon_{xx}=698\%$  **d.** $\varepsilon_{xx}=968\%$

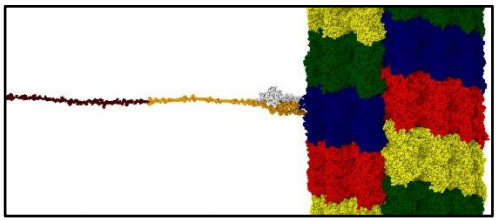
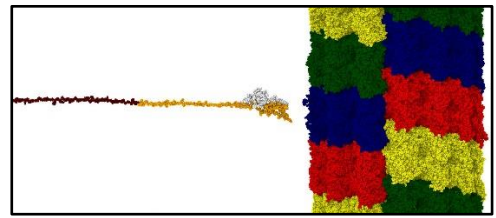

e. $\varepsilon_{xx}$=1108%    f. $\varepsilon_{xx}$=1128%

**Figure 6:** Observation during the pulling of single tau towards the -x direction (strain rate: $2\times10^9$ s$^{-1}$) away from MT surface. a. Initial stage (strain: 0%), b. unfolding of tau projection domain (strain: 409%), c. stretching of tau projection domain (strain: 698%), d. stretching of MT binding region (strain: 968%), e. onset of separation (strain: 1108%), f. after complete separation (strain: 1128%). Color legends: Red, blue, green and yellow: repeating helical units of MT, maroon: projection domain of tau, orange: MT binding site atoms of tau (including the interrepeats), white: N terminus tail of tau.

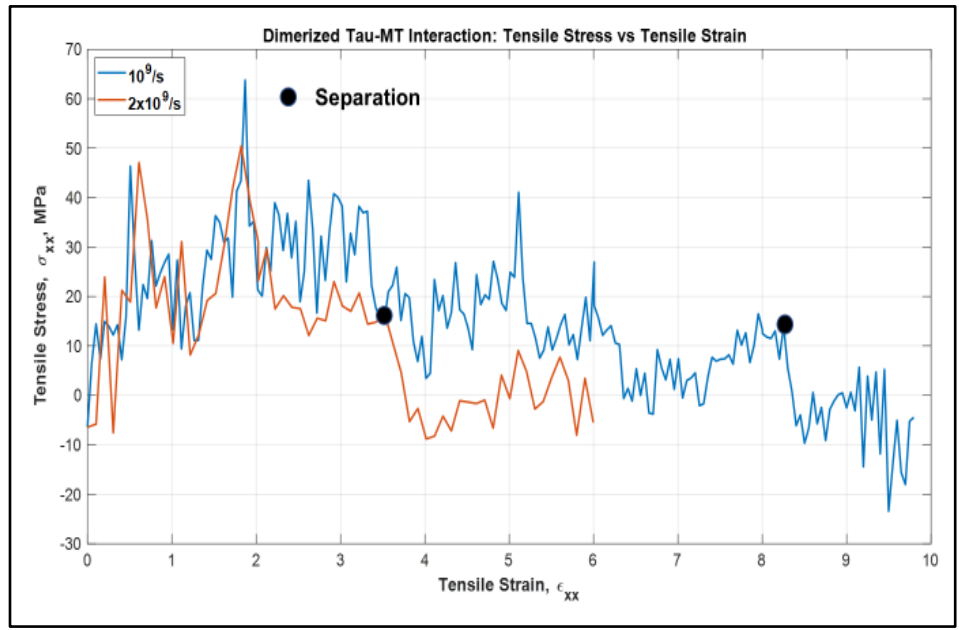

a

b

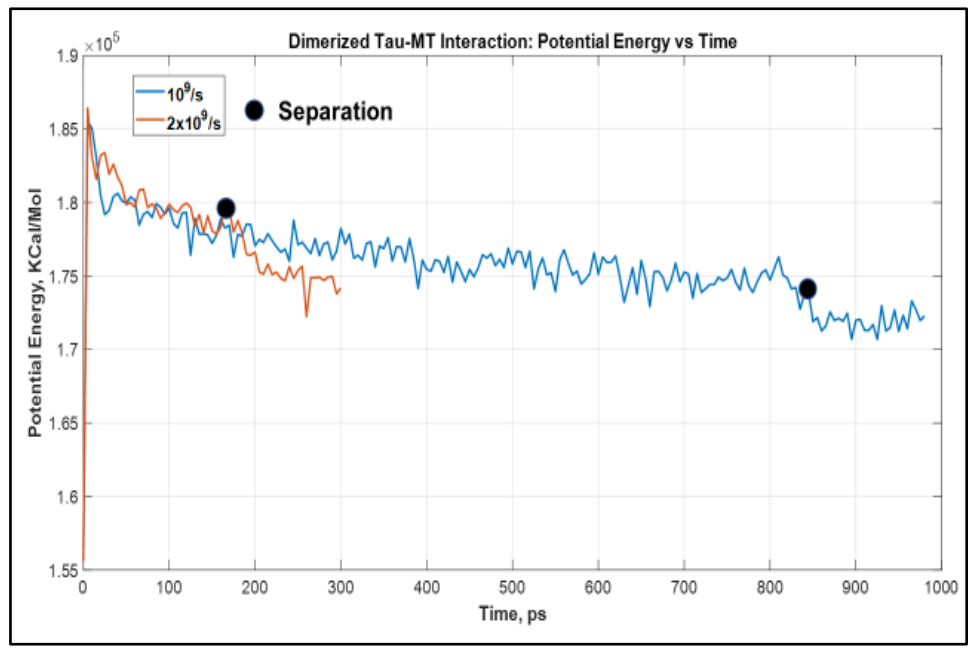

**Figure 7:** a. Stress vs Strain graph for the projection domain of tau during the pulling at different strain rates. The separation occurs early for higher strain rate (~360%), and much later for lower strain rate (~825%). b. Potential energy vs time plot for the dimerized tau-MT system. As expected, potential energy drastically reduces at separation (~360% strain for $2x10^9$ s$^{-1}$, ~825% strain for $1x10^9$ s$^{-1}$). For both cases, tau-tau bond is stronger than tau-MT bond.

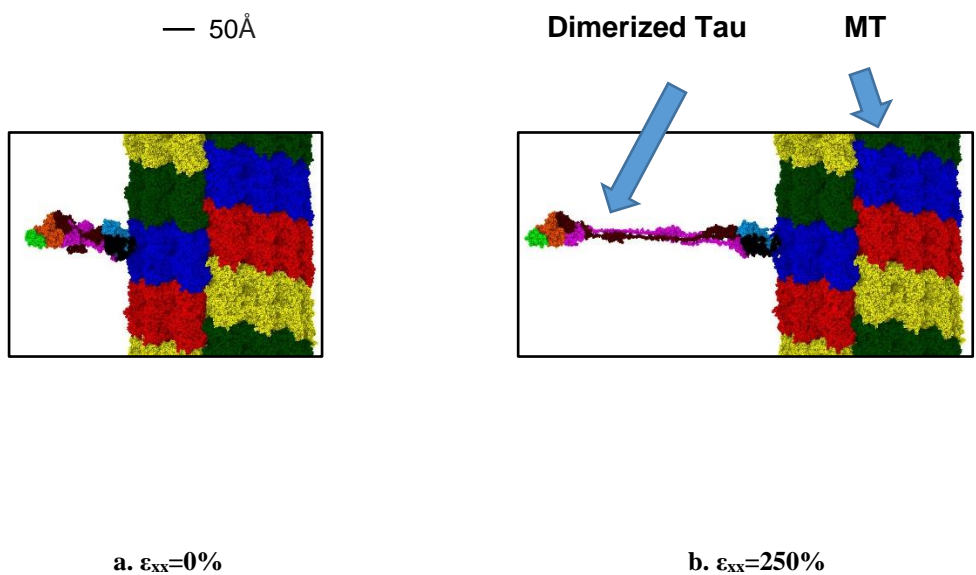

a. $\varepsilon_{xx}=0\%$          b. $\varepsilon_{xx}=250\%$

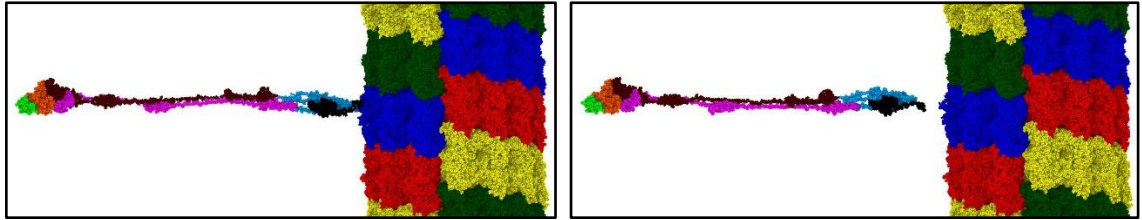

**c.** $\varepsilon_{xx}=350\%$  **d.** $\varepsilon_{xx}=354\%$

**Figure 8:** Observation during the dimerized tau pulling away from the MT surface (-x direction, strain rate: $2\times10^9$ s$^{-1}$). a. Initial stage (strain: 0%), b. unfolding and stretching of tau (strain: 250%), c. onset of separation (strain: 350%), d. after separation (strain: ~354%). Color legends: Red, blue, green and yellow: repeating helical units of MT, maroon: projection domain of tau 1, orange: MT binding site atoms of tau 1(including the interrepeats), light green: tail domain of tau 1, purple: projection domain of tau 2, light blue: MT binding site atoms of tau 2(including the interrepeats), black: tail domain of tau 2.

## 4. Discussion

In this study, we have analyzed the response of tau protein and tau-MT interaction from a strictly mechanical point of view. We have performed tensile tests on a predicted structure of tau protein to determine the single tau projection domain stiffness, dimerized tau separation stretch, tau-MT separation stretch, and comparison between tau-tau vs tau-MT bond. For a disordered protein, the confidence score (C-score) of -0.03 has been assumed as reliable in our simulation. The detail of quantification of the reliability for a protein structure predicted by i-TASSER (Zhang, 2008) is discussed in the supplementary material of this manuscript.

Admittedly, the strain rates that we have applied fall into high to very high range, but it facilitates us to obtain an insight of sub-axonal level response of this particular neural cytoskeletal component. In reality, the tissue level loading might be lower than the cellular level stress, meaning moderate level blow on the head may lead to high level tissue deformation, which eventually leads to extreme level of stress and failure in sub-axonal level components, supported by recent finite element method (FEM) studies on axon (Cloots et al., 2011, 2010). These studies show that axonal level anisotropy and cellular level heterogeneity might play instrumental role to determine failure criteria of the components, and injury level. Also, this level of strain rate is justified in the scenario of cavitation bubble collapse or blast wave exposure, which leads to intensely high stress in sub-axonal component (Wu and Adnan, 2018).

Earlier studies which modeled MT response under mechanical loading did not incorporate tau protein with detail mechanical properties, rather studied MT protofibril response where tau protein is considered as a viscoelastic spring (Ahmadzadeh et al., 2014), the properties of which were adapted from earlier characterization of pro- and anti-aggregant conformations of tau protein obtained by single molecule force spectroscopy (SMFS) (Wegmann et al., 2011). However, this study characterizes the response of only mutant conformation of the "repeat domain of tau", not the projection domain which is susceptible to unfolding and stretching under the application of high strain rate. Therefore, the current study is an effective extension of previous SMFS study with more comprehensive insight. In general, SMFS studies show the stretch of a molecule from force vs displacement perspective focusing on the detachment peak force at the contour length of the molecule used, but in TBI scenario, a more relevant representation is using directional stress vs strain which can perfectly differentiate between the unfolding and stretching region of a protein under pure mechanical loading.

For dimerized tau, we have re-generated the dimerized tau structure with overlapped projection domains depicted in the study of Rosenberg et al (Rosenberg et al., 2008). This study particularly established the importance of projection domain which determines the inter-MT distance in axonal bundle, as the tau-tau interaction is dependent on the length of projection domain, and interaction with a surface can be adhesive to repulsive (or a combination of both) based on the configuration. This study highlighted on the tau-tau interaction and tau-mica interaction, and substantiated that the interaction force is a function of projection domain length. The current study undertakes the missing aspect of the study: effect of shear that the dimerized conformation is susceptible to when undergoing through a high mechanical stress. Essentially our study shows that the projection domains of dimerized tau proteins are strong against shear and sliding force, and that the scenario it undergoes in strictly mechanical loading is highly dynamic, consisting both unfolding and stretching of both proteins. It further fortifies the observation that proline-rich region and N terminus combination, which is the projection domain, is highly stretchable while in the dimerized conformation, although the individual tau proteins might reach the failure region earlier. This specific observation shows that mere electrostatic and van der Waals bonds between the negatively charged N terminal region of one tau protein and positively charged prolin-rich region of another tau protein are sufficient to withstand mechanical loading to significant extent. This mechanical behavior of tau is also highlighted in the MT modeling work performed earlier, where one of the key prediction is that tau protein may elongate differently, and according to the position along the MT bundle (Ahmadzadeh et al., 2015, 2014). This high ability of tau protein to stretch aligns with another prediction of the study, that is, tau elongation facilitates the sliding of MT. The increased rigidity of the prolin-rich region can also be held responsible for the particular behavior of tau projection domain as suggested by NMR spectroscopy studies (Mukrasch et al., 2007).

The tau-MT model in the current study is developed as per the proposal of Chau et al, which shows specific MT binding sites on tau can facilitate bonding with tau binding sites on MT (Chau et al., 1998; Nogales et al., 1998). The proposal suggests that tau-MT interaction is highly dynamic, and that one MT binding region can interact with one or both subunits of MT (α and β). In our model, the tau protein has been attached to the specific location of the helically arranged protofibril to ensure that tau interacts with the C-terminal site by only R1 or R1-R2 interrepeat, and with the internal site by the rest of the MT binding sites (R2-R4), which was the scheme suggested by Chau et al. This is a relevant representation of a TBI scenario, where the sub-axonal level stretch on the cytoskeletal component is high for a significant timespan, but tau-MT bond is sufficiently strong to withstand mechanical load as long as the entire projection domain is not stretched. As stretching does not occur before unfolding of all the conformed portions of the projection domain, the MT structure instability is not invoked before the directional stress is developed in the MT binding region. The "jaw interaction" between the flanking domain of tau protein and acidic outside of MT surface is also responsible for such strong affinity, as seen between the projection domain and MT (partially), tail domain and MT (fully), suggesting that tau-MT bond is strong irrespective of the intervention of MT binding sites, which is proposed by earlier NMR studies (Mukrasch et al., 2007). However, in our case, both the intervention of flanking region and MT binding region are present, which facilitates stronger bond, as suggested by separate study (Preuss et al., 1997). Similar to dimerized tau system, this set of simulations shows that the interaction between charged portions (MT binding sites of tau and C-terminal as well as last 1/3$^{rd}$ portion of the C-terminal region excluding the terminal 12 amino acids) provides significant mechanical strength, even when susceptible to high to very high strain rate.

Finally, the dimerized tau-MT interaction system was a representation of more comprehensive scenario of TBI, in which there is a competition between the mechanical strength of tau-tau and tau-MT bond. As all of the involved regions in this case are charged (prolin-rich region of tau, negatively charged N-terminus region of tau, negatively charged outside surface of MT, etc.), we can assume that electrostatic interaction is more important in this case than the van der Waals bonds. The dimerized tau-MT bond shows that it requires development of more stress in tau-tau interaction region than required in tau-MT region, which suggests that prolin-rich region interaction with negatively charged projection domain is stronger than negatively charged MT surface interaction with positively charged MT binding sites, and under high strain rate, we can expect tau separation from MT surface. In case of multiple occasion of tau separation from MT surface in a single injury phenomenon, it may lead to the MT system collapse before the tau system collapse, as tau actually folds into stable conformation upon binding with MT (Kadavath et al., 2015).

As tau protein contains prolin-rich region, it is highly relevant to study its characteristics and assesses its behavior from neurodegenerative disease perspective. The importance of such prolin-rich proteins have been evident in separate neurodegenerative disorder studies (Gladkevich et al., 2007; Sochocka et al., 2019). These proteins are heavily implicated in neurodegenerative diseases and traumatic brain injury, which tau is also involved in. The PRR is a speculative binding site in proteins, so future studies could include protein biochemistry experiments focused on protein-protein interactions.

Evident from the strength of the tau-MT bond demonstrated in this paper, despite having a fast binding kinetics, we suggest that the intrinsic disorder of tau facilitates this phenomenon. Tau regularly alters its conformation, so its inherent flexibility is a likely source of the protein's ability to remain bound despite increased strain rate. We suggest tau absorbs the strain throughout its length and relies on the strong capability of the projection domain to remain bound to the microtubule during events where the brain undergoes significant strain and stretch. The strength of the tau-MT bond is particularly important for the field of traumatic brain injury, where strain and axonal stretch is thought to be a primary mechanism of injury in traumatic brain injury (Chung et al., 2005). Future studies might include studying comparing the IDP-substrate bond of other IDPs to see if the high bond strength is a common feature across IDPs as compared with ordered proteins.

Lastly, it is also evident from our MD simulation study that incorporation of physical chemistry perspective (such as posttranslational modification like domain focused or residue focused phosphorylation) along with the mechanical viewpoint is important to obtain comprehensive insight on tau protein behavior. Also, in all the tensile tests, tau protein has shown the dependence on the applied strain rate as single tau behaves as a stiffer material at higher strain rate in both unfolding and stretching region and dimerized tau separation and tau-MT separation stretches have shown strong dependency on strain rate, which suggests the importance of a separate study of viscoelastic characterization of tau protein.

## 5. Conclusion

In this paper, we have computationally determined the stiffness of projection domain of single tau protein and dimerized tau proteins, and the strength of tau-MT interface. The necessity of this study has been depicted by recent review study that pointed out that insight on high strain rate behavior of tau protein is currently absent in the literature. Due to the length scale associated with axonal cytoskeleton, MD simulation approach has been utilized as a viable maneuver. From our MD simulations, the major findings can be summarized as below:

1. Single tau protein is highly stretchable. It shows unfolding to a great extent before being purely stretched. The unfolding stiffness range is between 50MPa and 500MPa, while stretching stiffness can be >2GPa. The stiffness in both regions increases with the increase of the strain rate.
2. Dimerized tau-tau bond is strong, and the dimer structure does not dissociate before being stretched at >7.5 times of the initial length.
3. Tau protein can be separated from MT only at very high stretch (>11 times of the initial length of tau), and tau-MT bond is stronger than the MT subunit bond. Also, from the dimerized tau-MT simulation, we have obtained that tau-tau bond is stronger than tau-MT bond. We can hypothesize about mechano-chemical events which can trigger MT-tau separation, which have the timescale(s) than MD simulation can capture.
4. Strain rate affects the separation stretch and developed stress in tau for dimerized tau model, while it only affects the separation stretch significantly for tau-MT model. Higher strain rate causes early separation, and vice versa.

Bottom-up computational modeling of axon requires insight on mechanical behavior of individual components, and therefore, this study provides required insight on the strain rate dependent mechanical behavior of individual tau protein as well as tau-MT interface interaction. In injury biomechanics area and especially in multiscale traumatic brain injury (TBI) studies, these findings will play instrumental role to determine damage criteria at sub-axonal level. This MD simulation study particularly finds out sub-axonal level response of axonal cytoskeletal components, which are relevant to TBI scenario, where nanoscale injury propagates (axonal damage, MT instability, tau unfolding and stretching, tau-MT separation) due to macroscale impact (head injury).

**Data Availability**

The simulation data are available from the corresponding author upon reasonable request.

**Acknowledgements**

This work has been funded by the Computational Cellular Biology of Blast (C2B2) program through the Office of Naval Research (ONR) (Award # N00014-18-1-2082- Dr. Timothy Bentley, Program Manager). The authors acknowledge the Texas Advanced Computing Center (TACC) at The University of Texas at Austin for providing HPC resources that have contributed to the research results reported within this paper. URL: http://www.tacc.utexas.edu.

**Author Contributions**

M.I.K. performed prediction of the tau structure, designed the CHARMM models, ran all the simulations, interpreted the results, and contributed to writing the manuscript. K.G. provided insight on high strain rate dependent mechanical behavior of IDPs. A.A. designed the study and contributed to revising the manuscript. FH and KAM contributed to the simulation setup and review of the manuscript.

**Competing Interests:** The authors declare no competing interests.